\documentclass[10pt,a4paper, twocolumn, journal,transmag]{IEEEtran}
\usepackage[utf8]{inputenc}
\usepackage[english]{babel}
\usepackage{amsmath}
\usepackage{amsfonts}
\usepackage{amssymb}
\usepackage{makeidx}
\usepackage{graphicx}
\usepackage{xcolor}
\usepackage{hyperref}
\usepackage{tikz}

\usetikzlibrary{shapes.geometric, arrows}
\tikzstyle{encoder} = [rectangle, rounded corners, minimum width=1em, minimum height=2em,text centered, draw=black, fill=orange!70]
\tikzstyle{decoder} = [rectangle, rounded corners, minimum width=3em, minimum height=1.5em,text centered, draw=black, fill=yellow!50]
\tikzstyle{arrow} = [->,>=stealth, very thick]
\tikzstyle{meta} = [rectangle, rounded corners, minimum width=3em, minimum height=2.5em,text centered, draw=black, fill=blue!40]
\tikzstyle{trainee} = [rectangle, rounded corners, minimum width=3em, minimum height=1.5em,text centered, draw=black, fill=gray!30]
\tikzstyle{gnode} = [circle, text centered, draw=black, fill=gray!30]

\author{%
    \IEEEauthorblockN{Vincent ROGER, Jérôme FARINAS and Julien PINQUIER}
    \IEEEauthorblockA{IRIT, Université de Toulouse, CNRS, Toulouse, France}
}

\title{Deep Neural Networks for Automatic Speech Processing: A Survey from Large Corpora to Limited Data}

\begin{document}
    \IEEEtitleabstractindextext{%
	\begin{abstract}
        Most state-of-the-art speech systems are using Deep Neural Networks (DNNs).
        Those systems require a large amount of data to be learned.
        Hence, learning state-of-the-art frameworks on under-resourced speech languages/problems is a difficult task.
        Problems could be the limited amount of data for impaired speech.
        Furthermore, acquiring more data and/or expertise is time-consuming and expensive.
        In this paper we position ourselves for the following speech processing tasks: Automatic Speech Recognition, speaker identification and emotion recognition.
        To assess the problem of limited data, we firstly investigate state-of-the-art Automatic Speech Recognition systems as it represents the hardest tasks (due to the large variability in each language).
        Next, we provide an overview of techniques and tasks requiring fewer data.
        In the last section we investigate few-shot techniques as we interpret under-resourced speech as a few-shot problem.
        In that sense we propose an overview of few-shot techniques and perspectives of using such techniques for the focused speech problems in this survey.
        It occurs that the reviewed techniques are not well adapted for large datasets.
        Nevertheless, some promising results from the literature encourage the usage of such techniques for speech processing.
	\end{abstract}

    }

	\maketitle
	\section{Introduction}
    \IEEEPARstart{A}{utomatic} speech processing systems drastically improved the past few years, especially Automatic Speech Recognition (ASR) systems.
    It is also the case for other speech processing tasks such as speaker identification, emotion classification, etc.
    This success was made possible by the large amount of annotated data available combined with the extensive use of deep learning techniques and the capacity of modern Graphics Processing Units.
    Some models are already deployed for everyday usage such as your personal assistants on your smartphones, your connected speakers and so on.

    Nevertheless, challenges remain for automatic speech processing systems.
    They lack robustness against large vocabulary in real-world environment: this includes noises, distance from the speaker, reverberations and other alterations.
    Some challenges, such as CHIME \cite{barkerFifthCHiMESpeech2018}, provide data to let the community try to handle some of these problems.
    It is being investigated to improve the generalization of modern models by avoiding the inclusion of other annotated data for every possible environment.

    State-Of-The-Art (SOTA) techniques for most speech tasks require large datasets.
    Indeed, with modern DNN speech processing systems, having more data usually imply better performances.
    The TED-LIUM 3 from \cite{hernandezTEDLIUMTwiceMuch2018} (with 452 hours) provide more than twice the data of the TED-LIUM 2 dataset.
    Doing so, they obtain better results by training their model on TED-LIUM 3 than training their model over TED-LIUM 2 data.
    This improvement in performance for ASR systems is also observed with the LibriSpeech dataset (from~\cite{panayotovLibrispeechASRCorpus2015}).
    V. Panayotov et al. obtain better results on the Wall Street Journal (WSJ) test set by training a model over LibriSpeech dataset (1000 hours) than training a model over the WSJ training set (82 hours) \cite{panayotovLibrispeechASRCorpus2015}.

    This phenomenon, of having more data imply better performances, is also observable with the VoxCeleb 2 dataset compare to the VoxCeleb dataset: \cite{chungVoxCeleb2DeepSpeaker2018} increase the number of sentences from 100,000 utterances to one million utterances and increase the number of identities from 1251 to 6112 compared to the previous version of VoxCeleb.
    Doing so, they obtain better performances compare to training their model with the previous VoxCeleb dataset.\\


    With under-resourced languages (such as~\cite{dekaSpeechCorporaResourced2018}) and/or tasks (pathological detection with speech signals), we lack large datasets.
    By under-resourced, we mean limited digital resources (limited acoustic and text corpora) and/or a lack of linguistic expertise.
    For a more precise definition and details of the problem you may look \cite{besacierAutomaticSpeechRecognition2014}.
    Non-conventional speech tasks such as disease detection (such as Parkinson, gravity of ENT cancer and others) using audio are examples of tasks under resourced.
    Train Deep Neural Network models in such context is a challenge for these under-resourced speech datasets.
    This is especially the case for large vocabulary tasks.
    M. Moore et al. showed that recent ASR systems are not well adapted for impaired speech \cite{mooreWhistleblowingASRsEvaluating2018} and  M. B. Mustafa et al. showed the difficulties to adapt such models with limited amount of data \cite{mustafaSeverityBasedAdaptationLimited2014}.
    Few-shot learning consists of training a model using $k$-shot (where shot means an example per class), where $k \geq 1$ and $k$ is a low number.
    Training an ASR system on a new language, adapting an ASR system on pathological speech or doing a speaker identification with few examples are still complicated tasks.
    We think that few-shot techniques may be useful to tackle these problems.

    This survey will be focused on how to learn Deep Neural Network (DNN) models under low resources for speech data with non-overlapping mono signals.
    Therefore, we will first review SOTA ASR techniques that use a large amount of data (section~\ref{sec:asr}).
    Then we will review techniques and speech tasks (speaker identification, emotion recognition) requiring fewer data than SOTA techniques (section~\ref{sec:fewerdata}).
    We will also look into pathological speech processing for ASR using adaptation techniques (subsection~\ref{sec:domtrans}).
    Finally, we will review few-shot techniques for audio (section~\ref{sec:fewshot}) which is the focus of this survey.

    \section{Automatic Speech Recognition Systems}\label{sec:asr}

    In this section, we will review SOTA ASR systems using multi-models and end-to-end models.
    Here, we are focused on mono speech sequences $\mathbf{x} = [x_1, x_2, \ldots, x_n]$ where $x_i$ can be speech features or audio samples.
    ASR systems consist in matching $\mathbf{x}$ into a sequence of words $y=[y_1, y_2, \ldots, y_u]$ (where $u \leq n$).
    The systems reviewed were evaluated using Word Error Rate (WER) measure.

    \subsection{Multi-models}
    A multi-model approach consists in solving a problem using multiple models.
    Those models are designed to solve either sub-tasks (related to the problem) and the targeted task.
    The minimum configuration is with two models (let say $f$ and $g$) to solve a given task.
    Classically for the ASR task we can first learn an acoustic model (a phoneme classifier or equivalent sound units), then learn on top of it a language model that output the desired sequence of words.
    Hence, we have:
    \begin{equation}
      \hat{\mathbf{y}} = f(g(\mathbf{x}))
    \end{equation}
  with $f$ being the language model and $g$ being the acoustic model.
  Both can be learned separately or conjointly.
    Usually, hybrid models are used as acoustic models.


    %
    %

    %
    %
    Hybrid models consist in using probabilistic models with deterministic ones.
    Probabilistic models involve randomness using random variables combined with trained parameters.
    Hence, every prediction is sightly different on a given example~$\mathbf{x}$.
    Gaussian Mixture Models (GMMs) are an example of such models.
    Deterministic models do not involve randomness and every prediction are the same given an input $\mathbf{x}$.
    DNNs are an example of such models.
    A popular and efficient hybrid model is the DNN-Hidden Markov Model (DNN-HMM).
    DNN-HMM consists in replacing the GMMs that estimate the probability density functions by DNNs.
    The DNNs can be learned as phone classifiers.
    They form the acoustic model.
    This acoustic model is combined with a Language Model (LM) that maps the phonemes into a sequence of words.
    C. Lüscher et al. used DNN-HMMs combined with a Language Model to obtain SOTA on LibriSpeech test-other set (official augmented test set) \cite{luscherRWTHASRSystems2019}.
    This model process MFCC computed on the audio signals.
    Their best LM approach consisted in the use of Transformer from~\cite{vaswaniAttentionAllYou2017}.
    Transformers are autoregressive models (depending on the previous outputs of the models) using soft attention mechanisms.
    Soft attention consists in determining a glimpse $g$ over all possible glimpses such as:
    \begin{equation}
    g = \sum_{g' \in x}g' Pr(g'|a)
    \end{equation}
	 with $x$ being the input data and $a$ the attention parameters.
    Their best hybrid model got a Word Error Rate (WER) of 5.7\% for the test-other set and a WER of 2.7\% for test-clean set.

    \subsection{End-to-end systems}

    In end-to-end approaches, the goal is to determine a model $f$ that can do the mapping:
    \begin{equation}
    \hat{\mathbf{y}} = f(\mathbf{x})
    \end{equation}
    It will be learned straightforward from the $\mathbf{x}$ to the desired $\mathbf{y}$.
    Only supervised methods can be end-to-end to solve the speech tasks we are focused on.

    In ASR systems,~\cite{kimImprovedVocalTract2019} got SOTA on LibriSpeech test-clean official set.
    Compared to~\cite{luscherRWTHASRSystems2019} they used Vocal Tract Length Perturbation as the input of their end-to-end model.
    C. Kim et al. model is based on the Encoder-Decoder architecture using stacked LSTM for the encoder and LSTM combined with soft attention for the decoder \cite{kimImprovedVocalTract2019}.
    They obtain a WER of 2.44\% on test-clean and a WER of 8.29\% on test-other.
    Those results are close to~\cite{luscherRWTHASRSystems2019} (best hybrid model results) and show that end-to-end approaches are competitive compared to multi-model approaches.


   	\section{Techniques and tasks requiring fewer data}\label{sec:fewerdata}

	Some techniques require fewer data than the techniques of the previous section.
    In this section we will enumerate the principal ways to leverage (to our best knowledge) the lack of large datasets like unimpaired speech.
    We will also look into tasks requiring fewer data (speaker identification and emotion recognition).
    We will not talk of semi-supervised techniques that use a large amount of unsupervised data.

    \subsection{Data augmentation}

    The first way to leverage the lack of data is to artificially augment the number of data.
    To do so, classic approach consists for example in adding noise or deformation.
    Such as in \cite{parkSpecAugmentSimpleData2019}.
    They obtain near SOTA on Librispeech (1000 hours from \cite{panayotovLibrispeechASRCorpus2015}) with an end-to-end models.
    Nevertheless, they obtain SOTA results on SwitchBoard (300 hours from \cite{godfreySWITCHBOARDTelephoneSpeech1992}) with a WER of 6.8\%/14.1\% on the Switchboard/CallHome portion using shallow fusion and their data augmentation.
    But theses are handcrafted augmentations and some of them require additional audios (like adding noise).

    Some other approaches use generative models to have new samples such as in~\cite{chatziagapiDataAugmentationUsing2019, jiaoSimulatingDysarthricSpeech2018}.
    A. Chatziagapi et al. used conditional Generative Adversarial Networks (GAN) to generate new samples~\cite{chatziagapiDataAugmentationUsing2019}.
    Conditioned GAN are GAN where we can control the mode of the generated samples.
    Doing so, they balanced their initial dataset and obtain better results.
    Y. Jiao et al. used Deep Convolutional GANs to generate dysarthric speech and improve their results~\cite{jiaoSimulatingDysarthricSpeech2018}.

    \subsection{Domain transposition}\label{sec:domtrans}

    Another way to leverage the lack of data is to use domain transposition to avoid complex domain, here is some recent examples on speech:

    \begin{itemize}

        \item K. Wang et al. used GAN to dereverberate speech signal \cite{wangInvestigatingGenerativeAdversarial2018}.
    In their work, the generator is used as a mapping function of reverberated signals into dereverberated speech signals.

        \item L.-W. Chen et al. do vocal conversion using GAN with a controller mapping impaired speech to a representation space $z$ \cite{chenGenerativeAdversarialNetworks2019}.
    $z$ is then the input of the generator that is used as a mapping function to have unimpaired speech signals.

        \item S. Zhao et al. used Cycle GAN (framework designed for domain transfer) as an audio enhancer \cite{zhaoMultiTaskMultiNetworkJointLearning2019}.
    Their resulting model is SOTA on Chime-4 dataset.
    \end{itemize}

    \subsection{Models requiring fewer parameters}\label{sec:fewerparam}

    Having fewer data disallow the use of many parameters for Neural Network models to avoid overfitting.
    This is why some techniques tried to have models requiring fewer parameters.
    Here, we highlight some recent techniques that we find interesting:

    \begin{itemize}

        \item The use of SincNet, from~\cite{ravanelliInterpretableConvolutionalFilters2018}, layers to replace classic 1D convolutions over raw audio.
        Here, instead of requiring $window\_size$ parameters (with $window\_size$ being the window size of the 1D convolution) per filter, we only need two parameters per filter for every window size.
        Theses two parameters represent in a way (not directly) the values of the bandwidth at high and low energy.

        \item The use of LightGRU (LiGRU),  from~\cite{ravanelliLightGatedRecurrent2018}, based on the Gated Recurrent Unit (GRU) framework.
        LiGRU is a simplification of the GRU framework given some assumption in audio.
        They removed the reset gate of the GRU and used the ReLU activation function (combined with the Batch Normalization) instead of the $\tanh$ activation function.

        \item The use of quaternions Neural Networks, from \cite{parcolletSpeechRecognitionQuaternion2018}, for speech processing.
        The quaternion formulation allows the fuse of 4 dimensions into one inducing a drastic reduction of required parameters in their experiments (near 4 times).
    \end{itemize}

    \subsection{Multi-task approach}\label{sec:multitasks}

\begin{figure}[ht]
        \centering
        \begin{tikzpicture}[node distance=4em]
            \node(pred1){prediction $t_1$};
            \node(pred2)[right of=pred1, xshift=2.8em]{prediction $t_2$};
            \node(dots)[right of=pred2, xshift=1.5em]{\ldots};
            \node(predn)[right of=dots, xshift=2em]{prediction $t_n$};
            \node(dec1)[decoder, below of=pred1]{Decoder$_1$};
            \node(dec2)[decoder, below of=pred2]{Decoder$_2$};
            \node(decdots)[decoder, below of=dots]{\ldots};
            \node(decn)[decoder, below of=predn]{Decoder$_n$};
            \node(encoder)[encoder, below of=dec2, xshift=2.8em]{Encoder};
            \node(input)[below of=encoder]{Input signal};
            \draw[arrow] (dec1) -- (pred1);
            \draw[arrow] (dec2) -- (pred2);
            \draw[arrow] (decn) -- (predn);
            \draw[arrow] (encoder.north) -- ++(0,0.3) -| (dec1.south);
            \draw[arrow] (encoder.north) -- ++(0,0.3) -| (dec2.south);
            \draw[arrow] (encoder.north) -- ++(0,0.3) -| (decn.south);
            \draw[arrow] (input) -- (encoder);
        \end{tikzpicture}
        \caption{Multi-task architecture illustration. The output of the encoder is given to each decoder to have the prediction for each $t_i$ task.}
    \label{fig:multiT}
	\end{figure}
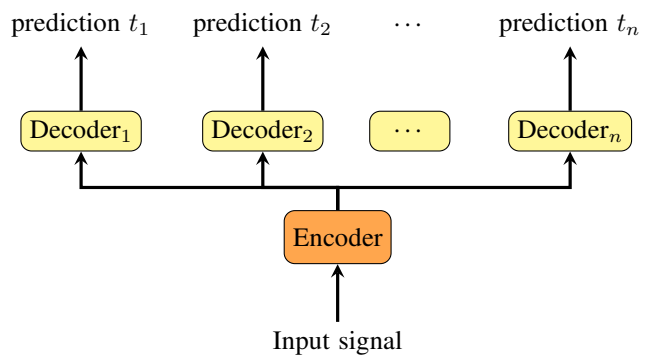

    Multi-task models can be viewed as an extension of the Encoder-Decoder architecture where you have a decoder per task with a shared encoder (like in Figure~\ref{fig:multiT}).
	Then those tasks are trained conjointly with classic feed-forward algorithms.
    The goal of a multi-task learning is to have an encoder outputting sufficient information for every task.
    Doing so, it can potentially improve the performances of each task compared to mono task architectures.
    It is a way to have a more representative encoder given the same amount of data.

    \label{emotion}
    In emotion recognition,~\cite{liImprovedEndtoEndSpeech2019} got SOTA results over a modified version of the IEMOCAP database to have a four-class problem.
	Those emotions are: angry, happy, neutral and sad.
	Y. Li et al. used an end-to-end multi-task system with only supervised tasks: gender identification and emotion identification \cite{liImprovedEndtoEndSpeech2019}.
    The resulting model achieve an overall accuracy for the emotion task (which is the main target) of 81.6\% and an average accuracy of each emotion category of 82.8\%.
    Using such approach allows them to achieve balanced results over unbalanced data.

    Nevertheless, using only supervised tasks requires multiple ground-truth for the targeted dataset.
    S. Pascual et al. used a combination of self-supervised tasks combined with unsupervised tasks to tackle this problem and used the resulting encoder for transfer learning \cite{pascualLearningProblemagnosticSpeech2019}.
    They recently improved this work in \cite{ravanelliMultitaskSelfsupervisedLearning2020} where they use more tasks, a recurrent unit on top of the encoder and denoising mechanisms using multiple data augmentation on their system.

    \subsection{Transfer Learning}\label{sec:transfer}
    Transfer learning techniques consist of using a pre-trained model and transfer its knowledge to solve a related problem/task.
    Usually we use the encoding part of the pre-trained model to initialize the model for the new problem/task.


    Contrastive Predictive Coding (CPC from~\cite{oordRepresentationLearningContrastive2018}) is an architecture to learn unsupervised audio representation using a 2-level architecture combined with a self-supervised loss.
    They achieved good results by transferring the obtained model for speaker identification and phone classification (on LibriSpeech dataset) compared to MFCC features.

    This work inspired~\cite{pascualLearningProblemagnosticSpeech2019}.
    They developed an unsupervised multi-task model (with certain losses being self-supervised) to obtain better encoders for transfer learning.
    They applied it on multiple tasks and obtain decent results on speaker identification (using VTCK), emotion recognition (using INTERFACE) and ASR (using TIMIT).

	The benefit of pre-trained network for transfer learning decrease as the target task diverges from the original task of the pre-trained network~\cite{yosinskiHowTransferableAre2014}.
    To tackle this, \cite{oordRepresentationLearningContrastive2018, pascualLearningProblemagnosticSpeech2019} attempt to have generic tasks with their unsupervised approach, and they obtained promising results.
    Also, the benefit of transfer learning decrease when the dissimilarity between the datasets increase~\cite{yosinskiHowTransferableAre2014}.
    This problem can discourage the use of transfer learning for some pathological speech.
    Whereas, Dysarthric and Accented Speech seems similar to speech in librispeech dataset according to \cite{shorPersonalizingASRDysarthric2019}.
    Where they successfully used transfer learning to improve their results over a 36.7 hours dataset.

    Nevertheless, \cite{mustafaSeverityBasedAdaptationLimited2014} showed that acoustic characteristics of unimpaired and impaired speech are very different.
    In the case of having few data such problems can be critical.
    It is why looking into few-shot techniques could be helpful.

	\section{Few-shot Learning and Speech}\label{sec:fewshot}

    In the previous sections, we reviewed models that require a large amount of data.
    This among of data is not always available such as for pathological speech.
    Google is trying to acquire more data of that nature\footnote{\url{https://blog.google/outreach-initiatives/accessibility/impaired-speech-recognition/}}.
    But acquiring such data can be quite expensive and time consuming.
    M. B. Mustafa et al. recommend the use of adaptive techniques to tackle limited amount of data problem in such case \cite{mustafaSeverityBasedAdaptationLimited2014}.
    But we think few-shot technique can be an other solution to this problem.
    Nevertheless, some non-common tasks such as pathological or dialect identification with few examples are still hard to train with SOTA techniques based on large speech datasets.
    This is why we investigate the following few-shot techniques and see the adaptations required for using them on speech datasets.

    \subsection{Few-shot Notations}

    Let consider a distribution $P$ from which we draw Independent Identically Distributed ($iid$) episodes $\mathcal{E}$, where
    $\mathcal{E}$ is composed of a support set $\mathcal{S}$, unlabeled data $\bar{\mathbf{x}}$ and a query set $\mathcal{Q}$.
    Support set correspond to the supervised samples the model has access to:
    \begin{equation}
    \mathcal{S} = \{(x_1, y_1), \ldots (x_s, y_s) \}
    \end{equation}
    with $x_i$ being samples and $y_i$ being the corresponding labels such as $y_i \in \{1, 2, \ldots, K\}$.
    $K$ being the number of classes appearing in $P$.
    The query set is composed of samples to classify $\hat{\mathbf{x}}$ with $\hat{\mathbf{y}}$ being the corresponding ground truth.\\

    To summarize, episodes drawn from $P$ have the following form:
    \begin{align}
        \begin{split}
            \mathcal{E} = \{ & \mathcal{S} =  \{(x_1, y_1), \ldots, (x_s, y_s) \},\\
            & \bar{\mathbf{x}} = (\bar{x}_1, \ldots, \bar{x}_r),\\
            & \mathcal{Q} = \{(\hat{x}_1, \hat{y}_1), \ldots, (\hat{x}_t, \hat{y}_t)\}
        \end{split}
    \end{align}
    with $s$, $r$ and $t$ fixed values that respectively represent the number of supervised samples for the support set, the number of unsupervised samples and the number of supervised samples for the query set.\\

    In this survey, we will focus on Few-Shot Learning techniques where $r=0$, $t\geq1$ and $s=kn$, with $n$ being the number of times each label appears for the support set and $k$ the number of classes selected from $P$, such as $k \leq K$.
    Hence, we have a $n$-shot with $k$ ways (or classes) for each episode.
    One-shot learning is just a special case of few-shot learning where $n = 1$.
    In some few-shot framework, we only sample one episode from $P$ and it represents our task.

    \subsection{Few-shot learning techniques}

    In this section we will review frameworks that impacted the few-shot learning field in image processing, frameworks with a formulation that seems adapted for speech processing and frameworks already successfully used by the speech community.\\

    \subsubsection{Siamese technique}
    Siamese Neural Networks are designed to be used per episode~\cite{kochSiameseNeuralNetworks2015}.
    They consist of measuring the distance between two samples and tell if they are similar or not.
    Hence, Siamese network uses the samples from the support set $\mathcal{S}$ as references for each class.
    It is then trained using all the combinations of samples from $\mathcal{S} \bigcup \mathcal{Q}$ which represent much more training than having only $s+t$ samples in classical feedforward frameworks.
    Siamese Networks take two samples ($x_1$ and $x_2$) as input and compute a distance between them, as follows:
    \begin{equation}
        \phi(x_1, x_2) = \sigma(\sum\boldsymbol{\alpha}|Enc(x_1) - Enc(x_2)|)
    \end{equation}
    with $Enc$ being a DNN encoder that represents the signal input, $\sigma$ being the sigmoid function, $\boldsymbol{\alpha}$ learnable parameters that weight the importance of each component of the encoder and $x_1$ and $x_2$ sampled from either the support set nor the queries set.\\

\begin{figure}[h]
        \centering
    \begin{tikzpicture}[thick]
        \node(data){$x_i$};
        \node(siam)[meta, right of=data, xshift=3.5em, minimum width=10em, minimum height=4em, font=\small, yshift=-1em, text depth = 5.5em]{Siamese Model};
        \node(model)[encoder, right of=data, xshift=1em]{$Enc$};
        \node(disc)[decoder, right of=model, xshift=2em, yshift=-1.3em]{$\phi$};
        \node(truth)[below of=data]{$\hat{x}_j$};
        \node(model2)[encoder, right of=truth, xshift=1em]{$Enc$};
        \node(prediction)[right of=disc, xshift=5em]{Same or Different};
        \draw [arrow] (data) -- (model);
        \draw [arrow] (truth) -- (model2);
        \draw [arrow] (model.east) -- ++ (0.3, 0) |- (disc.170);
        \draw [arrow] (model2.east) -- ++ (0.3, 0) |- (disc.190);
        \draw [arrow] (disc) -- (prediction);
    \end{tikzpicture}
    \caption{Example of comparison between a reference ($x_i$) and a new example ($\hat{x}_j$) from the query set.
        Where $Enc$ is the same network applied to both $x_i$ and $\hat{x}_j$.
    The model output the distance between $x_i$ and $\hat{x}_j$ class.}
    \label{fig:siam}
    \end{figure}
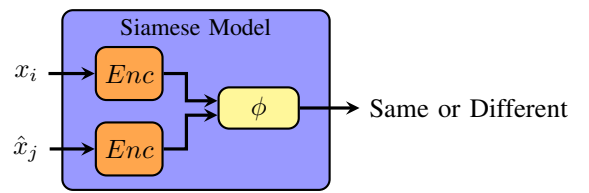

    To define the class of a new sample from $\mathcal{Q}$ or any new data, we have to compute the distance between each reference from $\mathcal{S}$ and the new sample.
    An example of comparison between a reference and a new example is shown in Figure~\ref{fig:siam}.
    Then, the class of the reference with the lowest distance become the prediction of the model.
    To learn such model, \cite{kochSiameseNeuralNetworks2015} used this loss function:
    \begin{align}
        \begin{split}
            \mathcal{L} = & \mathbb{E}_{y(x_i) = y(\tilde{x}_j)} \log(\phi(x_i, \tilde{x}_j)) +\\
                          & \mathbb{E}_{y(x_i) \neq y(\tilde{x}_j) } \log(1 - \phi(x_i, \tilde{x}_j))
        \end{split}
    \end{align}
    with $\mathbf{\tilde{x}} = [x_1, \ldots, x_s, \hat{x}_1, \ldots, \hat{x}_t]$ from $\mathcal{S}$ and $\mathcal{Q}$. $y(x)$ is a function that returns the label corresponding to the example $x$.
    Also, $\phi$ last layer should be a softmax.\\

    R. Eloff et al. used a modified version of this framework for Multimodal Learning (framework that is out of scope for this survey) between speech and image signal \cite{eloffMultimodalOneshotLearning2019}.
    The speech signals used consist of 11-digit number (zero to nine and oh) with the corresponding 10 images (oh and zero give the same images).
    The problem is to associate speech signals with the corresponding image.
    In their experiment, the model shows some invariances to speakers (accuracy of 70.12\% $\pm$ 0.68) using only a one-shot configuration, which is promising results.

    Siamese Neural Networks are not well adapted when the number of classes $K$ or the number of shots $q$ become too high.
    It increases the number of references to compare and the computation time to forward the model.
    It is mostly a problem for learning the model.
    After the model is learned, we can pre-calculate all representations for the support set to reduce this effect.
    Also, it drastically increases the number of combinations to do for training, this can be viewed as a positive point as we can truncate the number of combinations to use for training the model.
    This framework seems not adapted for end-to-end ASR with large vocabulary such as in the English speech (around 470,000 words).
    Maybe it will be sufficient for languages such as Esperanto language (around 16,780 words).
	The other way to use such a framework in ASR systems is to use it in hybrid models as an acoustic model.
    Where we can learn it on every phoneme (for example 44 phonemes/sounds in English) or more refined sound units. 

    Siamese framework seems interesting for tasks such as speaker identification.
    Indeed, this framework allows adding new speaker without retraining the model (supposing the model had generalized) or change the architecture of the model.
    We have to at least add one example of the new speaker to the references.
    Furthermore, Siamese formulation seems well adapted for speaker verification.
    Indeed, by replacing the pair $(\mathbf{x}, speaker\_id)$ by the pair $(\mathbf{x}, \mathcal{S}_{top5})$ we can do speaker verification with such technique.
    Where $\mathcal{S}_{top5}$ is a support set composed of signals from the 5 top predictions of the identification sub-task.

    Nevertheless, this framework will be limited if the number of speakers to identify become too high.
    Even so, it is possible to use such techniques in an end-to-end ASR system when the vocabulary is limited, such as in~\cite{eloffMultimodalOneshotLearning2019} experiment.\\

    \subsubsection{Matching Network}

    Matching Networks from~\cite{vinyalsMatchingNetworksOne2016} is a few-shot framework designed to be trained on multiple episodes.
    This framework is composed of one model~$\varphi$.
    This model is trained over a set of training episodes (with typically 5 to 25 ways).
    This model evaluates new examples given the support set $\mathcal{S}$ like in the Siamese framework:
    \begin{equation}
        \varphi(\hat{x}, \mathcal{S}) :\rightarrow \hat{y}
    \end{equation}

    In matching learning, $\varphi$ is as follows:
    \begin{equation}
        \varphi(\hat{x}, \mathcal{S}) = \sum_{(x_i, y_i) \in \mathcal{S}} a(\hat{x}, x_i) y_i
    \end{equation}
    with, $a$ being the attention kernel.\\

    In~\cite{vinyalsMatchingNetworksOne2016} this attention kernel is as follows:
    \begin{equation}
        a(\hat{x}, x_i) = \text{softmax}(c(f(\hat{x}), g(x_i)))
    \end{equation}
    where $c$ is the cosine distance, $f$ and $g$ are embedding functions.\\

    O. Vinyals et al. used a recurrent architecture to modulate the representation of $f$ using the support set $S$ \cite{vinyalsMatchingNetworksOne2016}.
    The goal is to have $f$ following the same type of representation of $g$.
    To do this, $g$ function is as follows:
    \begin{equation}
        g(x_i) = \overrightarrow{h_i} + \overleftarrow{h_i} + g'(x_i)
    \end{equation}
    where $\overrightarrow{h_i}$ and $\overleftarrow{h_i}$ represent a bi-LSTM output over $g'(x_i)$ which is a DNN\@.\\

    $f$ function is as follows:
    \begin{equation}
        f(\hat{x}) = attLSTM(f'(\hat{x}), g(S), m)
    \end{equation}
    with, $attLSTM$ being an LSTM with a fixed number of recurrences to do (here $m$), $g(S)$ represents the application of $g$ to each $x_i$ from the $S$ set.
    $f'$ is a DNN with the same architecture as $g'$, but not necessarily share the parameter values.
    Hence, training this framework consists in the maximization of the log likelihood of $\varphi$ given the parameters of $g$ and $f$.\\

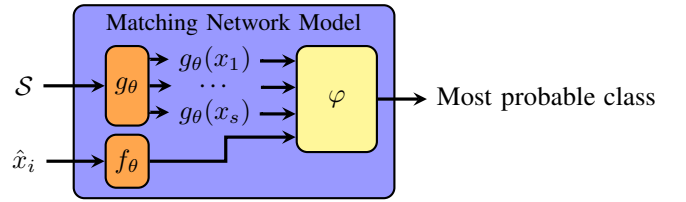
\begin{figure}[h]
        \centering
    \begin{tikzpicture}[thick]
        \node(data){$\mathcal{S}$};
        \node(siam)[meta, right of=data, xshift=5em, minimum width=12em, minimum height=4.5em, font=\small, yshift=-.6em, text depth = 6em]{Matching Network Model};
        \node(model)[encoder, minimum height=3em, right of=data, xshift=1em]{$g_\theta$};
        \node(disc)[decoder, minimum height=4em, right of=model, xshift=5em, yshift=-.5em]{$\varphi$};
        \node(truth)[below of=data]{$\hat{x}_i$};
        \node(model2)[encoder, right of=truth, xshift=1em]{$f_\theta$};
        \node(prediction)[right of=disc, xshift=5em]{Most probable class};
        \node(g1)[right of=model, yshift=1em, xshift=1ex]{$g_\theta(x_1)$};
        \node(dots)[right of=model, xshift=1ex]{\ldots};
        \node(gs)[right of=model, yshift=-1em, xshift=1ex]{$g_\theta(x_s)$};
        \draw [arrow] (data) -- (model);
        \draw [arrow] (truth) -- (model2);
        \draw [arrow] (model.east) ++ (0, 0.35) |- (g1.west);
        \draw [arrow] (model.east) -- ++ (0.29, 0);
        \draw [arrow] (model.east) ++ (0, -0.35) |- (gs.west);
        \draw [arrow] (model2.east) -- ++ (1, 0) |- (disc.223);
        \draw [arrow] (gs.east) |- (disc.200);
        \draw [arrow] (dots.east) ++ (0.28, 0) |- (disc.164);
        \draw [arrow] (g1.east) |- (disc.137);
        \draw [arrow] (disc) -- (prediction);
    \end{tikzpicture}
    \caption{Illustration of the Matching Network model to predict class of a new example $\hat{x}_i$.}
    \label{fig:match}
    \end{figure}

    Figure~\ref{fig:match} illustrates forward time of the Matching Network model.
    For forward time on new samples $g(\mathcal{S})$ can be pre-calculated to gain computation time.
    Nevertheless, as for Siamese networks, Matching networks have the same disadvantages when $q$ and/or $K$ become too high.
    Furthermore, adding new classes to a trained Matching Network model is not as easy as for Siamese Network models.
    Indeed, it requires retraining the Matching Network model to add an element to the support set.
    Whereas, Matching learning showed better results than the Siamese framework on image datasets from~\cite{vinyalsMatchingNetworksOne2016} experiments.
    It is why it should be investigated in speech processing to see if it is still the case.\\

    \subsubsection{Prototypical Networks}

    Prototypical Networks~\cite{snellPrototypicalNetworksFewshot2017} are designed to work with multiple episodes.
    In the prototypical framework, the model $\varphi$ does its predictions given the support set $\mathcal{S}$ of an episode such as the previously seen frameworks.
    This framework uses training episodes as mini-batches to obtain the final model.
    This model is formulated as follows:
    \begin{equation}
        \varphi(\hat{x}, S) = softmax_k(-d(f(\hat{x}), \mathbf{c}_k))
    \end{equation}
    where $\mathbf{c}_k$ is the prototype of the class $k$, $d$ being a Bregman divergence (for their useful properties in optimization, see \cite{snellPrototypicalNetworksFewshot2017} for more details) that also follow this property: $\mathbf{R}^n \times \mathbf{R}^n \rightarrow [0, +\inf[$.\\

    J. Snell et al. used the Euclidean distance for $d$ instead of the cosine distance used in Meta Learning and Matching Learning papers \cite{snellPrototypicalNetworksFewshot2017}.
    Doing so, they obtain better results in their experiments.
    Next, they go further by reducing the Euclidean to a linear function.

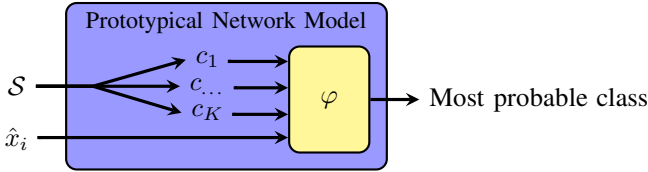
\begin{figure}[h]
        \centering
    \begin{tikzpicture}[thick]
        \node(data){$\mathcal{S}$};
        \node(siam)[meta, right of=data, xshift=5em, minimum width=12em, minimum height=4.5em, font=\small, yshift=0em, text depth = 5em]{Prototypical Network Model};
        \node(disc)[decoder, minimum height=4em, right of=model, xshift=5em, yshift=-.5em]{$\varphi$};
        \node(truth)[below of=data, yshift=0.85em]{$\hat{x}_i$};
        \node(prediction)[right of=disc, xshift=5em]{Most probable class};
        \node(g1)[right of=model, yshift=1em, xshift=1ex]{$c_1$};
        \node(dots)[right of=model, xshift=1ex]{$c_\text{\ldots}$};
        \node(gs)[right of=model, yshift=-1em, xshift=1ex]{$c_K$};
        \draw [arrow] (data) -- ++ (1, 0) -- (g1.west);
        \draw [arrow] (data) -- (dots);
        \draw [arrow] (data) -- ++ (1, 0) -- (gs.west);
        \draw [arrow] (truth.east) |- (disc.223);
        \draw [arrow] (gs.east) |- (disc.200);
        \draw [arrow] (dots.east) |- (disc.164);
        \draw [arrow] (g1.east) |- (disc.137);
        \draw [arrow] (disc) -- (prediction);
    \end{tikzpicture}
    \caption{Illustration of the Prototypical Network model to predict class of a new example $\hat{x}_i$.}
    \label{fig:proto}
    \end{figure}

    In the prototypical framework, there is only one prototype for each class~$k$ as illustred in Figure.~\ref{fig:proto}. It is computed such as:
    \begin{equation}
        \mathbf{c}_k = \frac{1}{|\mathcal{S}_k|} \sum_{(x_i, y_i) \in \mathcal{S}_k}f(x_i)
    \end{equation}
    with $f$ being a mapping function such as $\mathbb{R}^D \rightarrow \mathbb{R}^M$ and $\mathcal{S}_k$ being the samples with $k$ of the support set.\\

    Compared to Siamese and Matching Learning Networks, prototypical networks require only one comparison per class and not $q$ per class for $q$-shot learning like in Siamese and Matching Learning Networks.
    It is why this framework is less subject to the high computation problem for prediction of new samples as it is only influenced by high $K$.
    It will certainly be insufficient for end to end ASR systems on English language but it is a step forward to it.\\

	\subsubsection{Meta-Learning}

    Meta-learning \cite{raviOPTIMIZATIONMODELFEWSHOT2017} are designed to be learned on multiple episodes (also called datasets).
    In this framework a trainee model ($\mathcal{T}$) with parameters $\theta^{\mathcal{T}}$ is trained for every episode from the start of every episode.
    It usually has a classic DNN architecture.
    The support set and the query set in the episodes are considered as the training set and the test set for the trainee model.

    Along with this trainee model, a second model is learned: the meta model ($\mathcal{M}$) with parameters $\theta^{\mathcal{M}}$.
    This meta model is the key of meta learning, it consists in monitoring the trainee model by updating $\theta^{\mathcal{T}}$ parameters.
    To learn this meta model, sampling $iid$ episodes from $P$ to form the meta-dataset ($\mathcal{D}$) is suggested in \cite{raviOPTIMIZATIONMODELFEWSHOT2017}.
    This meta-dataset is composed of a training set ($\mathcal{D}_{train}$), a validation set ($\mathcal{D}_{valid}$) and a testing set ($\mathcal{D}_{test}$).

\begin{figure}[h]
        \centering
        \begin{tikzpicture}[node distance=3em]
            \node(itm1){$i_{t-1}$};
            \node(ftm1)[below of=itm1, yshift=2em]{$f_{t-1}$};
            \node(theta)[below of=ftm1, yshift=-2em]{$\theta_{t-1}^\mathcal{T}$};
            \node(meta)[meta, right of=itm1, yshift=-0.5em, xshift=2em]{$\mathcal{M}$};
            \node(tout1)[below of=meta, yshift=-0.5em]{$[\mathcal{L}_t^\mathcal{T}, \nabla_{\theta_{t-1}^\mathcal{T}}\mathcal{L}_t^\mathcal{T}, \theta_{t-1}^\mathcal{T}]$};
            \node(trainee)[trainee, below of=tout1]{$\mathcal{T}_{t-1}$};
            \node(ej1)[below of=trainee]{$\mathcal{E}_j$};
            \node(meta1)[meta, right of=meta, xshift=10em]{$\mathcal{M}$};
            \node(tout2)[below of=meta1, yshift=-0.5em]{$[\mathcal{L}_{t+1}^\mathcal{T}, \nabla_{\theta_{t}^\mathcal{T}}\mathcal{L}_{t+1}^\mathcal{T}, \theta_{t}^\mathcal{T}]$};
            \node(trainee1)[trainee, below of=tout2]{$\mathcal{T}_{t}$};
            \node(ej2)[below of=trainee1]{$\mathcal{E}_j$};
            \node(etc1)[right of=meta1, xshift=2em]{\ldots};
            \node(etc2)[right of=trainee1, xshift=2em]{\ldots};

            \draw[transform canvas={yshift=0.75ex}, ->] (meta) -- (meta1) node[above,midway] {$i_t$};
            \draw[transform canvas={yshift=-0.75ex}, ->] (meta) -- (meta1) node[below,midway] {$f_t$};
            \draw[->] (meta) -- (trainee1) node[midway, below] {$\theta_t^\mathcal{T}$};
            \draw[->] (tout1) -- (meta);
            \draw[->] (tout2) -- (meta1);
            \draw[->] (trainee) -- (tout1);
            \draw[->] (trainee1) -- (tout2);
            \draw[->] (itm1) -- ++(3.5em, 0);
            \draw[->] (ftm1) -- ++(3.5em, 0);
            \draw[->] (theta) -- (trainee);
            \draw[->] (ej1) -- (trainee);
            \draw[->] (ej2) -- (trainee1);
        \end{tikzpicture}
        \caption{Meta-Learning illustration for training over episode $\mathcal{E}_j$ at step $t$. Here the Meta model $\mathcal{M}$ process the different training step of the trainee $\mathcal{T}$ as a sequence.}
        \label{fig:meta}
    \end{figure}
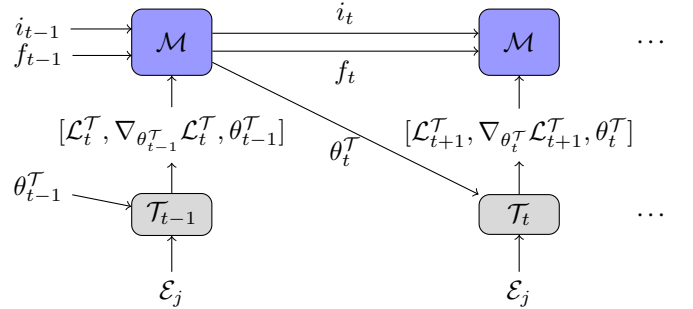

    While the trainee model is training on an episode $\mathcal{E}_j$, the meta model is charged to update its parameters:
    \begin{equation}
        \theta^{\mathcal{T}_j}_t = \mathcal{M}(\theta^{\mathcal{T}_j}_{t-1}, \mathcal{L}^{\mathcal{T}_j}, \nabla_{\theta^{\mathcal{T}_j}_{t-1}}\mathcal{L}^{\mathcal{T}_j})
    \end{equation}
    with $\mathcal{L}^{\mathcal{T}_j}$ being the loss function of the trainee model learned over the episode $\mathcal{E}_j$ and $\theta_{t-1}^{\mathcal{T}_j}$ are the parameters of the trainee model at step $t-1$.
    Also, $\mathcal{M}$ has to guess initial weights of the trainee models at step $t=0$ ($\theta^{\mathcal{T}_j}_0$).
    The learning curve (loss) of the trainee model over $\mathcal{E}_j$ is viewed in \cite{raviOPTIMIZATIONMODELFEWSHOT2017} as a sequence that can be the input of the meta model $\mathcal{M}$.
    For simplicity, we will use the notation of $\mathcal{T}$ instead of $\mathcal{T}_j$ for the next paragraphs.
    Figure \ref{fig:meta} illustrate the learning steps of the trainee using the meta model.

    \paragraph{Trainee parameters update}
    S. Ravi and H. Larochelle identify the learning process of $\mathcal{T}$ using classic feedforward update on episode $E_j$ to be similar with the $c_t$ update gate of the LSTM framework \cite{raviOPTIMIZATIONMODELFEWSHOT2017}.
    In the meta learning framework, the update gate $c_t$ of the LSTM framework is then used as the $\theta^\mathcal{T}_t$ estimator, such as:
    \begin{equation}\label{eq:lstmTrain}
        \theta_t^\mathcal{T} = f_t \odot \theta_{t-1}^\mathcal{T} + i_t \odot \tilde{\theta_t^\mathcal{T}}
    \end{equation}

    with $\tilde{\theta_t^\mathcal{T}} = -\alpha_t \nabla_{\theta_{t-1}^\mathcal{T}} \mathcal{L}_t^\mathcal{T}$ being the update term of the parameters $\theta_{t-1}^\mathcal{T}$, $f_t$ being the forget gate and $i_t$ the update gate.

    \paragraph{Parameters of the meta model}
    Both $i_t$ and $f_t$ are part of the Meta learner.
    In the meta-learning framework, the update gate is formulated as follows:
    \begin{equation}
        i_t = \sigma(\mathbf{W}_I.[\nabla_{\theta_{t-1}^\mathcal{T}}\mathcal{L}_t^\mathcal{T}, \mathcal{L}_t^\mathcal{T}, \theta_{t-1}^\mathcal{T}, i_{t-1}] + \mathbf{b}_I)
    \end{equation}
    with the $\mathbf{W}_I$ and $\mathbf{b}_I$ being parameters of $\mathcal{M}$.
    The update gate is used to control update term in~\ref{eq:lstmTrain} like the learning rate in classic feedforward approach.

    Next, the forget gate in the meta-learning framework is formulated as follows:
    \begin{equation}
        f_t = \sigma(\mathbf{W}_F.[\nabla_{\theta_{t-1}^\mathcal{T}}\mathcal{L}_t^\mathcal{T}, \mathcal{L}_t^\mathcal{T}, \theta_{t-1}^\mathcal{T}, f_{t-1}] + \mathbf{b}_F)
    \end{equation}
      with $\mathbf{W}_F$ and $\mathbf{b}_F$ parameters of $\mathcal{M}$.
      This gate is here to decide whether the learning of the trainee should restart or not.
      This can be useful to get out of a sub-optimal local minimum.
      Note that this gate is not present in classic feedforward approaches (where this gate is equal to one).\\

    The trainee model ($\mathcal{T}$) of this framework can be any kind of model such as a Siamese Neural Network.
    Hence, it can have the advantages of this framework.
    It also can avoid the Siamese neural network disadvantages as it can use any other framework (usually classic DNN).
    This framework is interesting for speech processing to learn efficient models (in terms of learning speed) when we have multiple ASR tasks with different vocabulary.
    For example, let say we have these kinds of speech episodes: dialing numbers, commands to a robot A and commands to a robot B. The model can initialize good filters for the first layers (as it is still speech processing).
    Another example could be learning acoustic models for multiple languages (with each episode corresponding to a language).\\

    \subsubsection{Graph neural network}

    The use of Graph Neural Network (GNN) is used by
    V. Garcia and J. Bruna introduce the use of Graph Neural Network (GNN) in their few-shot framework \cite{garciaFewShotLearningGraph2018}.
    This framework is designed to be used with multiple episodes they called tasks.
    In this framework, one model is used over a complete graph $G$.
    $G = (V, E)$ where every node corresponds to an example.
    GNN for few-shot learning consists in applying Graph Convolutions Layers over the graph $G$.

    Initial vertices construction to guess the ground truth of a query $\tilde{x_i}$ from the query set $\mathcal{Q}$:
    \begin{equation}
        \begin{split}
            V^{(0)} = ( & (Enc(x_1), h(y_1)), \ldots, (Enc(x_s), h(y_s)), \\
                        & (Enc(\bar{x_1}), u), \ldots, (Enc(\bar{x_r}), u)\\
                        & (Enc(\tilde{x_i}), u))
        \end{split}
    \end{equation}
    where $Enc$ is an embedding extraction function (a Neural Network or any classic feature extraction technique), $h$ the one-hot encoding function and $u=K^{-1} \mathbf{1}_K$ an uniform distribution for examples with unknown labels (the unsupervised ones from $\bar{\mathbf{x}}$ and/or from the query set $\mathcal{Q}$).\\

    From now the vertices at each layer $l$ (with 0 being the initial vertices) will be denoted:
    \begin{equation}
    V^{(l)} = (v_1, \ldots, v_n)
    \end{equation}
    where $n = s+r+1$ and $V^{(l)} \in \mathbb{R}^{n*d_l}$.\\

    Every layers in GNN are computed as follows:
    \begin{equation}
        V^{(l+1)} = Gc(V^{(l)}, A^{(l)})
    \end{equation}
    with $A^{(l)}$ being the adjacency operators constructed from $V^{(l)}$ and $Gc$ being the graph convolution.

    \paragraph{The adjacency operators construction}

    The adjacency operator us a set:
    \begin{equation}
        A^{(l)} = \{\tilde{A}^{(l)}, \mathbf{1}\}
    \end{equation}
    with $\tilde{A}^{(l)}$ being the adjacency matrix of $V^{(l)}$.\\

    For every $(i, j) \in E$ (recall we have complete graphs), we compute the values of the adjacency matrix such as:
    \begin{equation}
         \tilde{A}^{(l)}_{i, j} = \phi(v_i^{(l)}, v_j^{(l)})
    \end{equation}

    where:
    \begin{equation}
        \phi(v_i^{(l)}, v_j^{(l)}) = f(|v_i^{(l)} - v_j^{(l)}|)
    \end{equation}
    with $f$ being a multi-layer perceptron with its parameter denoted $\theta_f$.
    $\tilde{A}^{(l)}$ is then normalized using the softmax function over each line.

    \begin{figure}[h]
        \centering
        \begin{tikzpicture}[node distance=4em]
            \node(v1)[gnode, align=center]{\small $v_i$};
            \node(v2)[gnode, below of=v1, align=center, yshift=-2em]{\small $v_j$};
            \node(v3)[gnode, right of=v1, xshift=3.5em, align=center]{\small $v_k$};
            \node(v4)[gnode, right of=v2, xshift=3.5em, align=center]{\small $v_u$};
            \node(vi)[left of=v1, xshift=0em, yshift=1.5em]{\scriptsize $v_i=[Enc(x_i), h(y_i)]$};
            \node(vj)[left of=v2, xshift=0em, yshift=-1.5em]{\scriptsize $v_j = [Enc(x_j), h(y_j)]$};
            \node(vk)[right of=v3, xshift=0em, yshift=1.5em]{\scriptsize $v_k = [Enc(x_k), h(y_k)]$};
            \node(vu)[right of=v4, xshift=0em, yshift=-1.5em]{\scriptsize $v_u = [Enc(\bar{x}), h(u)]$};
            \draw (v1) -- (v2) node[left,midway]{\small$A_{i, j}^{(0)}$};
            \draw (v1) -- (v3) node[above,midway]{\small$A_{i, k}^{(0)}$};
            \draw (v1) -- (v4) node[above, midway, xshift=-2.3em, yshift=1.8em, right]{\small$A_{i, u}^{(0)}$};
            \draw (v2) -- (v3) node[below, midway, xshift=-2.5em, yshift=-1.9em, right]{\small$A_{j, k}^{(0)}$};
            \draw (v2) -- (v4) node[below,midway]{\small$A_{j, u}^{(0)}$};
            \draw (v3) -- (v4) node[right,midway]{\small$A_{k, u}^{(0)}$};
        \end{tikzpicture}
        \caption{Illustration of the input of the first layer (or Graph Convolution) of a GNN. Here we have three samples (represented by vertices $v_i$, $v_j$ and $v_k$) in the support set and one query (represented by the vertex $v_u$).}
        \label{fig:GNN}
    \end{figure}
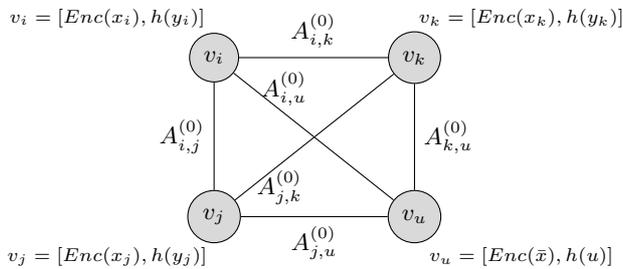

    \paragraph{Graph convolution}

    The graph convolution requires the construction of the adjacency operators set and is computed as follows:
    \begin{equation}
        Gc(V^{(l)}, A^{(l)}) = \rho(\sum_{B \in A} BV^{(l)} \theta^{(k)}_{B, l})
    \end{equation}

    with $B$ being an adjacency operator from $A$, $\theta_{B, l}^{(k)} \in \mathbb{R}^{d_{l-1}, d_l}$ learnable parameters and $\rho$ being a point wise linearity (usually leaky ReLU).

    \paragraph{Training the model}

    The output of the resulting GNN model is a mapping of the vertices to a $K$-simplex that give the probability of $\tilde{x_i}$ being in class $k$.
    V. Garcia and J. Bruna used the cross-entropy to learn the model other all examples in the query set $\mathcal{Q}$ \cite{garciaFewShotLearningGraph2018}.
    Hence, the GNN few-shot framework consists in learning $\theta_f$ and $\theta_{1, l} \ldots \theta_{card(A), l}$ parameters over all episodes.

    \paragraph{Few-shot GNN on audio}

    This framework was used by \cite{zhangFewShotAudioClassification2019} on 5-way audio classification problems.
    The 5 ways episodes are randomly selected from the initial dataset: AudioSet~\cite{gemmekeAudioSetOntology2017} for creating the 5-ways training episodes and \cite{zhangFastSVMTraining2006} data to create the 5-ways test episodes.

    S. Zhang et al. compare the use of per class attention (or intra-class) and global attention which gave the best results \cite{zhangFewShotAudioClassification2019}.
    They applied it for each layer.
    Their experiments were done for 1-shot, 5-shots and 10-shots with the respective accuracy of $69.4\% \pm 0.66$, $78.3\% \pm 0.46$ and $83.6\% \pm 0.98$.
    Such results really motivate us in the path of few-shot learning for speech signals.
    Nevertheless, this framework does not allow the use of many classes and shots per episode which increase the number of nodes and thus the computations in forward time.
    Hence, it is not suited for large vocabulary problems.

    \section{Summary and Future directions}\label{sec:conclu}

	In this survey, we investigated few-shot techniques for speech usage.
    In order to do so, we started with state-of-the-art speech processing systems.
    These systems require a large amount of data and are not suited for under-resourced speech problems.
    We also looked into techniques requiring fewer data using data augmentation, domain transposition, models requiring fewer parameters, multi-task approach and transfer learning.
    Nevertheless, these techniques are less efficient in a data-limited context.
    Next, we studied few-shot techniques and how well the different frameworks are adapted for classical speech tasks.

    The main drawback of the reviewed techniques is the amount of computation required for large datasets (like LibriSpeech from \cite{panayotovLibrispeechASRCorpus2015}) compared to SOTA models we reviewed in section~\ref{sec:asr}.
    Nevertheless, we considered some recent works already using few-shot techniques on speech with promising results.
    Such techniques seem useful for classical speech tasks on impaired speakers.
    Moreover, we think it can be useful for unconventional speech tasks like measuring the intelligibility of a person (with impaired or unimpaired speakers) to help the re-education process (by identifying the problems faster).
    Acquiring a large amount of data is painful for some patients (with severe pathologies).
    We believe that few-shot techniques may help the community to tackle this problem.
    To see the interest of such techniques we will work on a benchmark for different speech tasks.
    We will do some adaptations when necessary, but we think that we can use the different frameworks straightforward.
    After that, we plan to use the technique with the best results on this benchmark as a base for learning the concept of intelligibility.

	\bibliography{survey.bib}

\begin{thebibliography}{10}
\providecommand{\url}[1]{#1}
\csname url@samestyle\endcsname
\providecommand{\newblock}{\relax}
\providecommand{\bibinfo}[2]{#2}
\providecommand{\BIBentrySTDinterwordspacing}{\spaceskip=0pt\relax}
\providecommand{\BIBentryALTinterwordstretchfactor}{4}
\providecommand{\BIBentryALTinterwordspacing}{\spaceskip=\fontdimen2\font plus
\BIBentryALTinterwordstretchfactor\fontdimen3\font minus
  \fontdimen4\font\relax}
\providecommand{\BIBforeignlanguage}[2]{{%
\expandafter\ifx\csname l@#1\endcsname\relax
\typeout{** WARNING: IEEEtran.bst: No hyphenation pattern has been}%
\typeout{** loaded for the language `#1'. Using the pattern for}%
\typeout{** the default language instead.}%
\else
\language=\csname l@#1\endcsname
\fi
#2}}
\providecommand{\BIBdecl}{\relax}
\BIBdecl

\bibitem{barkerFifthCHiMESpeech2018}
J.~Barker, S.~Watanabe, E.~Vincent, and J.~Trmal, ``\BIBforeignlanguage{en}{The
  {{Fifth}} '{{CHiME}}' {{Speech Separation}} and {{Recognition Challenge}}:
  {{Dataset}}, {{Task}} and {{Baselines}}},'' in
  \emph{\BIBforeignlanguage{en}{Interspeech 2018}}.\hskip 1em plus 0.5em minus
  0.4em\relax {ISCA}, Sep. 2018, pp. 1561--1565.

\bibitem{hernandezTEDLIUMTwiceMuch2018}
F.~Hernandez, V.~Nguyen, S.~Ghannay, N.~Tomashenko, and Y.~Est{\`e}ve,
  ``{{TED}}-{{LIUM}} 3: Twice as much data and corpus repartition for
  experiments on speaker adaptation,'' in \emph{Speech and {{Computer}} - 20th
  {{International Conference}}}, vol. 11096, Sep. 2018, pp. 198--208.

\bibitem{panayotovLibrispeechASRCorpus2015}
V.~Panayotov, G.~Chen, D.~Povey, and S.~Khudanpur,
  ``\BIBforeignlanguage{en}{Librispeech: {{An ASR}} corpus based on public
  domain audio books},'' in \emph{\BIBforeignlanguage{en}{2015 {{IEEE
  International Conference}} on {{Acoustics}}, {{Speech}} and {{Signal
  Processing}} ({{ICASSP}})}}.\hskip 1em plus 0.5em minus 0.4em\relax {South
  Brisbane, Queensland, Australia}: {IEEE}, Apr. 2015, pp. 5206--5210.

\bibitem{chungVoxCeleb2DeepSpeaker2018}
J.~S. Chung, A.~Nagrani, and A.~Zisserman,
  ``\BIBforeignlanguage{en}{{{VoxCeleb2}}: {{Deep Speaker Recognition}}},'' in
  \emph{\BIBforeignlanguage{en}{Interspeech 2018}}.\hskip 1em plus 0.5em minus
  0.4em\relax {ISCA}, Sep. 2018, pp. 1086--1090.

\bibitem{dekaSpeechCorporaResourced2018}
B.~Deka, J.~Chakraborty, A.~Dey, S.~Nath, P.~Sarmah, S.~R. Nirmala, and
  S.~Vijaya, ``Speech corpora of under resourced languages of north-east
  india,'' in \emph{2018 Oriental {{COCOSDA}} - International Conference on
  Speech Database and Assessments, Miyazaki, Japan, May 7-8, 2018}, 2018, pp.
  72--77.

\bibitem{besacierAutomaticSpeechRecognition2014}
L.~Besacier, E.~Barnard, A.~Karpov, and T.~Schultz,
  ``\BIBforeignlanguage{en}{Automatic speech recognition for under-resourced
  languages: {{A}} survey},'' \emph{\BIBforeignlanguage{en}{Speech
  Communication}}, vol.~56, pp. 85--100, Jan. 2014.

\bibitem{mooreWhistleblowingASRsEvaluating2018}
M.~Moore, H.~Venkateswara, and S.~Panchanathan,
  ``\BIBforeignlanguage{en}{Whistle-blowing {{ASRs}}: {{Evaluating}} the
  {{Need}} for {{More Inclusive Speech Recognition Systems}}},'' in
  \emph{\BIBforeignlanguage{en}{Interspeech 2018}}.\hskip 1em plus 0.5em minus
  0.4em\relax {ISCA}, Sep. 2018, pp. 466--470.

\bibitem{mustafaSeverityBasedAdaptationLimited2014}
M.~B. Mustafa, S.~S. Salim, N.~Mohamed, B.~{Al-Qatab}, and C.~E. Siong,
  ``\BIBforeignlanguage{en}{Severity-{{Based Adaptation}} with {{Limited Data}}
  for {{ASR}} to {{Aid Dysarthric Speakers}}},''
  \emph{\BIBforeignlanguage{en}{PLoS ONE}}, vol.~9, no.~1, p. e86285, Jan.
  2014.

\bibitem{luscherRWTHASRSystems2019}
C.~L{\"u}scher, E.~Beck, K.~Irie, M.~Kitza, W.~Michel, A.~Zeyer,
  R.~Schl{\"u}ter, and H.~Ney, ``\BIBforeignlanguage{en}{{{RWTH ASR Systems}}
  for {{LibriSpeech}}: {{Hybrid}} vs {{Attention}}},'' in
  \emph{\BIBforeignlanguage{en}{Interspeech 2019}}.\hskip 1em plus 0.5em minus
  0.4em\relax {ISCA}, Sep. 2019, pp. 231--235.

\bibitem{vaswaniAttentionAllYou2017}
A.~Vaswani, N.~Shazeer, N.~Parmar, J.~Uszkoreit, L.~Jones, A.~N. Gomez,
  {\L}.~Kaiser, and I.~Polosukhin, ``Attention is {{All}} you {{Need}},'' in
  \emph{Advances in {{Neural Information Processing Systems}} 30}, I.~Guyon,
  U.~V. Luxburg, S.~Bengio, H.~Wallach, R.~Fergus, S.~Vishwanathan, and
  R.~Garnett, Eds.\hskip 1em plus 0.5em minus 0.4em\relax {Curran Associates,
  Inc.}, 2017, pp. 5998--6008.

\bibitem{kimImprovedVocalTract2019}
C.~Kim, M.~Shin, A.~Garg, and D.~Gowda, ``\BIBforeignlanguage{en}{Improved
  {{Vocal Tract Length Perturbation}} for a {{State}}-of-the-{{Art
  End}}-to-{{End Speech Recognition System}}},'' in
  \emph{\BIBforeignlanguage{en}{Interspeech 2019}}.\hskip 1em plus 0.5em minus
  0.4em\relax {ISCA}, Sep. 2019, pp. 739--743.

\bibitem{parkSpecAugmentSimpleData2019}
D.~S. Park, W.~Chan, Y.~Zhang, C.-C. Chiu, B.~Zoph, E.~D. Cubuk, and Q.~V. Le,
  ``{{SpecAugment}}: {{A Simple Data Augmentation Method}} for {{Automatic
  Speech Recognition}},'' \emph{Interspeech 2019}, pp. 2613--2617, Sep. 2019.

\bibitem{godfreySWITCHBOARDTelephoneSpeech1992}
J.~J. Godfrey, E.~C. Holliman, and J.~McDaniel, ``{{SWITCHBOARD}}:
  {{Telephone}} speech corpus for research and development,'' in
  \emph{[{{Proceedings}}] {{ICASSP}}-92: 1992 {{IEEE International Conference}}
  on {{Acoustics}}, {{Speech}}, and {{Signal Processing}}}, vol.~1.\hskip 1em
  plus 0.5em minus 0.4em\relax {IEEE}, 1992, pp. 517--520.

\bibitem{chatziagapiDataAugmentationUsing2019}
A.~Chatziagapi, G.~Paraskevopoulos, D.~Sgouropoulos, G.~Pantazopoulos,
  M.~Nikandrou, T.~Giannakopoulos, A.~Katsamanis, A.~Potamianos, and
  S.~Narayanan, ``\BIBforeignlanguage{en}{Data {{Augmentation Using GANs}} for
  {{Speech Emotion Recognition}}},'' in
  \emph{\BIBforeignlanguage{en}{Interspeech 2019}}.\hskip 1em plus 0.5em minus
  0.4em\relax {ISCA}, Sep. 2019, pp. 171--175.

\bibitem{jiaoSimulatingDysarthricSpeech2018}
Y.~Jiao, M.~Tu, V.~Berisha, and J.~Liss, ``Simulating dysarthric speech for
  training data augmentation in clinical speech applications,'' in \emph{{{IEEE
  International Conference}} on {{Acoustics}}, {{Speech}} and {{Signal
  Processing}}}, Apr. 2018.

\bibitem{wangInvestigatingGenerativeAdversarial2018}
K.~Wang, J.~Zhang, S.~Sun, Y.~Wang, F.~Xiang, and L.~Xie,
  ``\BIBforeignlanguage{en}{Investigating {{Generative Adversarial Networks}}
  based {{Speech Dereverberation}} for {{Robust Speech Recognition}}},''
  \emph{\BIBforeignlanguage{en}{Interspeech 2018}}, pp. 1581--1585, Sep. 2018.

\bibitem{chenGenerativeAdversarialNetworks2019}
L.-W. Chen, H.-Y. Lee, and Y.~Tsao, ``\BIBforeignlanguage{en}{Generative
  {{Adversarial Networks}} for {{Unpaired Voice Transformation}} on {{Impaired
  Speech}}},'' in \emph{\BIBforeignlanguage{en}{Interspeech 2019}}.\hskip 1em
  plus 0.5em minus 0.4em\relax {ISCA}, Sep. 2019, pp. 719--723.

\bibitem{zhaoMultiTaskMultiNetworkJointLearning2019}
S.~Zhao, C.~Ni, R.~Tong, and B.~Ma, ``\BIBforeignlanguage{en}{Multi-{{Task
  Multi}}-{{Network Joint}}-{{Learning}} of {{Deep Residual Networks}} and
  {{Cycle}}-{{Consistency Generative Adversarial Networks}} for {{Robust Speech
  Recognition}}},'' in \emph{\BIBforeignlanguage{en}{Interspeech 2019}}.\hskip
  1em plus 0.5em minus 0.4em\relax {ISCA}, Sep. 2019, pp. 1238--1242.

\bibitem{ravanelliInterpretableConvolutionalFilters2018}
M.~Ravanelli and Y.~Bengio, ``\BIBforeignlanguage{en}{Interpretable
  {{Convolutional Filters}} with {{SincNet}}},'' in
  \emph{\BIBforeignlanguage{en}{{{NIPS}} 2018 {{Workshop IRASL}}}}, Nov. 2018.

\bibitem{ravanelliLightGatedRecurrent2018}
M.~Ravanelli, P.~Brakel, M.~Omologo, and Y.~Bengio,
  ``\BIBforeignlanguage{en}{Light {{Gated Recurrent Units}} for {{Speech
  Recognition}}},'' \emph{\BIBforeignlanguage{en}{IEEE Transactions on Emerging
  Topics in Computational Intelligence}}, vol.~2, no.~2, pp. 92--102, Apr.
  2018.

\bibitem{parcolletSpeechRecognitionQuaternion2018}
T.~Parcollet, M.~Ravanelli, M.~Morchid, G.~Linar{\`e}s, and R.~De~Mori,
  ``Speech recognition with quaternion neural networks,'' in \emph{{{NeurIPS}}
  2018 - {{IRASL}}}, Nov. 2018.

\bibitem{liImprovedEndtoEndSpeech2019}
Y.~Li, T.~Zhao, and T.~Kawahara, ``\BIBforeignlanguage{en}{Improved
  {{End}}-to-{{End Speech Emotion Recognition Using Self Attention Mechanism}}
  and {{Multitask Learning}}},'' in \emph{\BIBforeignlanguage{en}{Interspeech
  2019}}.\hskip 1em plus 0.5em minus 0.4em\relax {ISCA}, Sep. 2019, pp.
  2803--2807.

\bibitem{pascualLearningProblemagnosticSpeech2019}
S.~Pascual, M.~Ravanelli, J.~Serr{\`a}, A.~Bonafonte, and Y.~Bengio,
  ``\BIBforeignlanguage{en}{Learning {{Problem}}-{{Agnostic Speech
  Representations}} from {{Multiple Self}}-{{Supervised Tasks}}},'' in
  \emph{\BIBforeignlanguage{en}{Interspeech 2019}}.\hskip 1em plus 0.5em minus
  0.4em\relax {ISCA}, Sep. 2019, pp. 161--165.

\bibitem{ravanelliMultitaskSelfsupervisedLearning2020}
M.~Ravanelli, J.~Zhong, S.~Pascual, P.~Swietojanski, J.~Monteiro, J.~Trmal, and
  Y.~Bengio, ``Multi-task self-supervised learning for {{Robust Speech
  Recognition}},'' \emph{arXiv:2001.09239 [cs, eess]}, Jan. 2020.

\bibitem{oordRepresentationLearningContrastive2018}
A.~van~den Oord, Y.~Li, and O.~Vinyals,
  ``\BIBforeignlanguage{en}{Representation {{Learning}} with {{Contrastive
  Predictive Coding}}},'' \emph{\BIBforeignlanguage{en}{CoRR}}, Aug. 2018.

\bibitem{yosinskiHowTransferableAre2014}
J.~Yosinski, J.~Clune, Y.~Bengio, and H.~Lipson, ``How transferable are
  features in deep neural networks?'' in \emph{Advances in {{Neural Information
  Processing Systems}} 27}, Z.~Ghahramani, M.~Welling, C.~Cortes, N.~D.
  Lawrence, and K.~Q. Weinberger, Eds.\hskip 1em plus 0.5em minus 0.4em\relax
  {Curran Associates, Inc.}, 2014, pp. 3320--3328.

\bibitem{shorPersonalizingASRDysarthric2019}
J.~Shor, D.~Emanuel, O.~Lang, O.~Tuval, M.~Brenner, J.~Cattiau, F.~Vieira,
  M.~McNally, T.~Charbonneau, M.~Nollstadt, A.~Hassidim, and Y.~Matias,
  ``\BIBforeignlanguage{en}{Personalizing {{ASR}} for {{Dysarthric}} and
  {{Accented Speech}} with {{Limited Data}}},'' in
  \emph{\BIBforeignlanguage{en}{Interspeech 2019}}.\hskip 1em plus 0.5em minus
  0.4em\relax {ISCA}, Sep. 2019, pp. 784--788.

\bibitem{kochSiameseNeuralNetworks2015}
G.~Koch, R.~Zemel, and R.~Salakhutdinov, ``\BIBforeignlanguage{en}{Siamese
  {{Neural Networks}} for {{One}}-shot {{Image Recognition}}},''
  \emph{\BIBforeignlanguage{en}{ICML Deep Learning Workshop}}, p.~8, 2015.

\bibitem{eloffMultimodalOneshotLearning2019}
R.~Eloff, H.~A. Engelbrecht, and H.~Kamper, ``Multimodal {{One}}-shot
  {{Learning}} of {{Speech}} and {{Images}},'' in \emph{{{ICASSP}} 2019 - 2019
  {{IEEE International Conference}} on {{Acoustics}}, {{Speech}} and {{Signal
  Processing}} ({{ICASSP}})}, May 2019, pp. 8623--8627.

\bibitem{vinyalsMatchingNetworksOne2016}
O.~Vinyals, C.~Blundell, T.~Lillicrap, k.~{kavukcuoglu}, and D.~Wierstra,
  ``Matching {{Networks}} for {{One Shot Learning}},'' in \emph{Advances in
  {{Neural Information Processing Systems}} 29}, D.~D. Lee, M.~Sugiyama, U.~V.
  Luxburg, I.~Guyon, and R.~Garnett, Eds.\hskip 1em plus 0.5em minus
  0.4em\relax {Curran Associates, Inc.}, 2016, pp. 3630--3638.

\bibitem{snellPrototypicalNetworksFewshot2017}
J.~Snell, K.~Swersky, and R.~Zemel, ``Prototypical {{Networks}} for
  {{Few}}-shot {{Learning}},'' in \emph{Advances in {{Neural Information
  Processing Systems}} 30}, I.~Guyon, U.~V. Luxburg, S.~Bengio, H.~Wallach,
  R.~Fergus, S.~Vishwanathan, and R.~Garnett, Eds.\hskip 1em plus 0.5em minus
  0.4em\relax {Curran Associates, Inc.}, 2017, pp. 4077--4087.

\bibitem{raviOPTIMIZATIONMODELFEWSHOT2017}
S.~Ravi and H.~Larochelle, ``\BIBforeignlanguage{en}{Optimization as a
  {{Model}} for {{Few}}-{{Shot Learning}}},'' in
  \emph{\BIBforeignlanguage{en}{{{ICLR}} 2017}}, 2017, p.~11.

\bibitem{garciaFewShotLearningGraph2018}
V.~Garcia and J.~Bruna, ``\BIBforeignlanguage{en}{Few-{{Shot Learning}} with
  {{Graph Neural Networks}}},'' in \emph{\BIBforeignlanguage{en}{{{ICLR}}
  2018}}, 2018, p.~13.

\bibitem{zhangFewShotAudioClassification2019}
S.~Zhang, Y.~Qin, K.~Sun, and Y.~Lin, ``\BIBforeignlanguage{en}{Few-{{Shot
  Audio Classification}} with {{Attentional Graph Neural Networks}}},'' in
  \emph{\BIBforeignlanguage{en}{Interspeech 2019}}.\hskip 1em plus 0.5em minus
  0.4em\relax {ISCA}, Sep. 2019, pp. 3649--3653.

\bibitem{gemmekeAudioSetOntology2017}
J.~F. Gemmeke, D.~P.~W. Ellis, D.~Freedman, A.~Jansen, W.~Lawrence, R.~C.
  Moore, M.~Plakal, and M.~Ritter, ``\BIBforeignlanguage{en}{Audio {{Set}}:
  {{An}} ontology and human-labeled dataset for audio events},'' in
  \emph{\BIBforeignlanguage{en}{2017 {{IEEE International Conference}} on
  {{Acoustics}}, {{Speech}} and {{Signal Processing}} ({{ICASSP}})}}.\hskip 1em
  plus 0.5em minus 0.4em\relax {New Orleans, LA}: {IEEE}, Mar. 2017, pp.
  776--780.

\bibitem{zhangFastSVMTraining2006}
S.~Zhang, H.~Jiang, S.~Zhang, and B.~Xu, ``\BIBforeignlanguage{en}{Fast {{SVM
  Training Based}} on the {{Choice}} of {{Effective Samples}} for {{Audio
  Classification}}},'' \emph{\BIBforeignlanguage{en}{INTERSPEECH 2006 -
  ICSLP}}, p.~4, 2006.

\end{thebibliography}
	\bibliographystyle{IEEEtran}

    \section*{Acknowledgment}
    Vincent Roger doctorate is founded by Federal University of Toulouse and Occitanie Region n\textsuperscript{o}2018-1290 (ALDOCT n\textsuperscript{o}500). This work is part of the ANR-18-CE45-0008 RUGBI project founded by French National Research Agency.

\end{document}